\documentstyle[psfig,rotating]{mn}

\newif\ifAMStwofonts
\AMStwofontstrue

\def\wp{{w_{+}}}
\def\wm{{w_-}}
\def\mup{{\mu_+}}
\def\mum{{\mu_-}}
\def\nup{{\nu_+}}
\def\num{{\nu_-}}
\def\lp{{\lambda_+}}
\def\lm{{\lambda_-}}

\def\Vp{{V_+}}
\def\Vm{{V_-}}
\def\gsim{~\rlap{$>$}{\lower 1.0ex\hbox{$\sim$}}}

\def\ltsim{\lower.5ex\hbox{$\; \buildrel < \over \sim \;$}}
\def\gtsim{\lower.5ex\hbox{$\; \buildrel > \over \sim \;$}}
\def\ltsim{\lower.5ex\hbox{$\; \buildrel < \over \sim \;$}}
\def\gtsim{\lower.5ex\hbox{$\; \buildrel > \over \sim \;$}}

\def\dd{\,{\rm d}}

\include{rotating}

\begin{document}
\textheight 9in
\title[Self-similar solutions]{General Relativistic Self-Similar Solutions in Cosmology}

\author[Nusser] { Adi Nusser \\\\
Physics Department- Technion, Haifa 32000, Israel}

\maketitle

\begin{abstract}
We present general relativistic solutions for  self-similar  spherical 
perturbations in an expanding cosmological background of cold pressure-less gas. We focus on solutions having  shock discontinuities propagating  in the surrounding  cold  gas. The pressure, $p$, and energy-density, $\mu$, in the shock-heated matter
are assumed to obey $p=w\mu$, where $w$ is a positive constant.   
Consistent solutions are found for shocks
propagating from the symmetry center of a region of a positive density excess over the background. In these solutions, shocks  exist outside the radius marking the event horizon of the black hole which would be present in a shock-less collapse. For large jumps in the energy-density at the
shock, a black hole is avoided altogether and the solutions are regular 
at the center. The shock-heated gas does not contain 
any sonic points, provided the motion of the cold gas ahead of the
shock deviates significantly from the Hubble flow. 
For shocks propagating in the uniform background, sonic points 
always appear for small jumps in the energy-density.
We also discuss self-similar solutions without shocks in fluids
with  $w<-1/3$.
\end{abstract}
 
                           
\section{introduction}

Detailed studies of generic self-gravitating  systems 
are only possible with the aid of complex numerical methods. 
Part of the complication arises from the long range nature of gravitational interactions.
Thus, even equilibrium states, when they exist, do not in general obey standard  statistical 
mechanics. This  is correct for both Newtonian gravity as well as general relativity (GR), but the latter adds its peculiar technical and conceptual challenges.    
Analytic or semi-analytic treatment of  systems possessing 
special symmetries are, therefore, of great interest. 
Gravity is the dominant force dictating the growth of structure in the universe. 
Thus, self-gravitating  perturbations in an expanding cosmological
has traditionally gained a great deal of attention (cf. Peebles 1990). 
Here we present a class of solutions which describe the evolution of spherical
perturbations in an expanding Einstein-de Sitter cosmological background. 
We focus on the evolution of self-similar  perturbations in GR. 
Self-similarity occurs when  dimensionless physical variables can be expressed as functions of  
some combination of the time and space coordinates. Self-similar systems offer 
great simplification since they are described by ordinary differential equations.
Newtonian treatment of cosmological self-similar perturbations  have been presented  in great detail previously in the literature  (e.g. Fillmore \& Goldreich 1984; Bertschinger 1985; Chuzhoy \& Nusser 2000). Nevertheless, it is instructive to review the general features of these
solutions. In cosmology, we assume that the density fluctuations
are extremely small at time, $t_i$, close to the  the Big Bang singularity.  
An isolated small density perturbation  having  the form  $\delta \mu \sim r^{-s}$ ($r$ is the distance from the symmetry center) at $t_i$, develops a self-similar behavior  at sufficiently
late times, $t\gg t_i$\footnote{An additional assumption which is often made is that the 
initial expansion rate of matter at  $r$ is equal to that of the unperturbed background. This assumption is, however, 
unnecessary  as long as the deviations from the background expansion rate are not great.}. 
Consider first a positive density perturbation in  a universe made solely of dust (i.e. collision-less matter). 
As time goes-by a shell of matter begins to expand at a rate lower than the background until 
it reaches a maximum distance from the center. The shell then falls inward
and ends up oscillating around the center together with shells which  have reached
it at earlier times. 
 According to  this picture, the radius, $r_1$, of the shell which is at maximum expansion 
 at time $t$ plays an important role.  
 For $r>r_1$,  matter is
 still  expanding, while at $r\ll r_1$ shells of matter are oscillating in
 the central  potential well generated by their self-gravity.   
The self-similar variable can then be taken to be 
the ratio, $r/r_1(t)$. 
For a universe filled with a collisional fluid (gas), self-similarity 
develops provided the initial pressure is negligible and that the
gas is able to shock and evolve adiabatically. In this case,
an inner region of hot shocked gas develops. The  shock bounding this region propagates  away from  the center at a rate $\dd r_1(t)/\dd t$.

Self-similarity in GR is obtained
for a more restricted class of initial perturbations than in Newtonian gravity. 
 GR admits self-similar solutions only for specific values of the power law index $s$, 
while Newtonian dynamics does not impose such restrictions.   
  Further, the pressure, $p$, must be related linearly to the energy-density, $\mu$,  of matter (Cahil \& Taub 1971).
Carr \& Hawking (1974) and Carr \& Yahil (hereafter CY90)  have studied   self-similar perturbations in an expanding universe. Their treatment, however, did not 
include shocks which might form as a result of collapse or energy injection into 
the system. Here we complement the work of CY90
 by exploring solutions which involve shock discontinuities. We assume that the 
 isotropic cosmic background is made of cold pressure-less matter that is able to shock.
We will  present solutions in which shocks propagate from a central point into the uniform cosmic background
or into the outer regions of a perturbation with  positive excess of energy-density in the cold gas.   
Shock-heated gas  will be assumed to obey the equation of state $p=w\mu$ ($w>0$).
We also present  solutions with no shocks for $w<-1/3$, a case which has not been 
discussed at length in the literature before.

The paper is organized as follows. In \S\ref{sec:eom} we summarize  GR equations 
for self-similar perturbations in comoving coordinates.
For a detailed  discussion of these equations we refer the reader to  CY90 
and Carr \& Coley (2000) (hereafter CC02). 
In \S\ref{sec:shock} we discuss the formation of shocks and the jump conditions
relating the physical variables on both sides of the shock discontinuity. 
In \S\ref{sec:num} we present full numerical solutions. We conclude with a general
discussion in \S\ref{sec:disc}.

\section{The  self-similar GR equations in spherical symmetry}
\label{sec:eom}

We work with comoving coordinates as they are particularly convenient  
for the study of cosmological perturbations.   With only minor changes of notation 
our description of the equations closely follows CY90 and CC02. 
 Denoting the time, the radius and the two spherical angular coordinates 
by $t$, $r$, $\theta$ and $\phi$, the space-time metric is expressed in terms of the 
interval $\dd s^2$ as (e.g. Landau \& Lifshitz 1975),
\begin{equation}
\dd s^2 =e^{2\nu}\dd t^2-e^{2 \lambda}\dd r^2 -R^2\dd \Omega^2\; ,
\end{equation}
where $\nu$, $\Lambda$ and $R$ are, in general, functions of $t$ and $r$, and
$\dd \Omega^2=\dd \theta^2+
\sin^2\theta\dd \phi^2$.
In comoving coordinates the energy-momentum tensor has the form 
$T^{\mu\nu}=(\mu+p)u^\mu u^\nu-pg^{\mu\nu}$ where 
$u^\mu=\dd t/\dd s=(e^{-\nu},0,0,0)$. For an equation of state $p=w \mu$ where 
$w$ is a constant, self-similar solutions in which 
 dimensionless variables are functions of  $z\equiv r/t$ alone can be obtained. 
We use the following dimensionless 
variables\footnote{The variable $\xi$ defined here is related to $x(z) $ 
in CY90 by $\xi^{1/w}$.},       
\begin{equation}
\label{eq:def}
S(z)\equiv \frac{R}{r}\; , \quad \xi(z)\equiv(\mu r^2)^{-1/(1+w)}\quad 
{\rm and} \quad
M(z)\equiv \frac{m}{R} \; ,
\end{equation}
where \begin{equation}
m(r,t)=\frac{R}{2}
\left[1+e^{-2\nu}\left(\frac{\partial R }{\partial t}\right)^2
-e^{-2\lambda}\left(\frac{\partial R }{\partial r}\right)^2
\right] \; .
\end{equation}
Using $T^{\alpha}_{\beta;\alpha}=0 $, gives 
$\partial_t \lambda+2\partial_t R/R=-2\partial_t \mu/(p+\mu)$
and $\partial_r\nu=-2\partial_r p/(p+\mu)$. For $p=w\mu$, these equations yield
\begin{equation}
e^\nu=\beta \xi^w z^{2 w/(1+w)}\quad {\rm and }\quad e^{-\lambda}= \gamma \xi^{-1} S^2 \; ,
\end{equation}
Substituting these relations in the remaining Einstein 
equations and using $\partial /\partial t= -(z/t)\dd/\dd z$
and $\partial /\partial r=(z/r)\dd/\dd z$  yield 
\begin{equation}
\label{eq:phys1}
\ddot S+\dot S+
\left[S+\left(1+w\right)\dot S\right]\left(\frac{2}{1+w}\frac{\dot S}{S}-\frac{\dot \xi}{\xi}\right)=0 \; ,
\end{equation}
\begin{eqnarray}
\nonumber
\left(\frac{2 w \gamma^2}{1+w}\right)S^4+\frac{2}{\beta^2}\frac{\dot S}{S}
\xi^{2-2w} z^{(2-2 w)/(1+w)}\\
-\left(V^2-w\right)\gamma S^4\frac{\dot \xi}{\xi}=(1+w)\xi^{1-w} \; ,
\label{eq:phys2}
\end{eqnarray}
\begin{equation}
\label{eq:phys3}
M=S^2\xi^{-(1+w)}\left[1+\left(1+w\right)\frac{\dot S}{S}\right]\; ,
\end{equation}
\begin{equation}
\label{eq:phys4}
M=\frac{1}{2}+\frac{1}{2\beta^2}\xi^{-2 w}z^{2(1-w)/(1+w)}-
\frac{1}{2}\gamma^2\xi^{-2}S^6\left(1+\frac{\dot S}{S}\right)\; ,
\end{equation}
and
\begin{equation}
V=(\beta\gamma)^{-1}\xi^{1-w}S^{-2}z^{(1-w)/(1+w)}\; ,
\end{equation}
where an over-dot denotes a derivative with respect to $y=\ln z$.

A particular solution of these equations describes the Einstein-e Sitter universe, i.e. 
the  flat critical density  Friedmann-Robertson-Walker (FRW) uniform density background. This solution is  
\begin{equation}
\xi=z^{-2/(1+w)} \quad , \quad S=z^{-2/(3+3w)}\; ,
\end{equation}
with 
\begin{equation}
\beta=\frac{\sqrt2}{\sqrt3(1+w)}\quad , \quad \gamma=\frac{3+3w}{1+3w}\; .
\end{equation}
The space-time interval is 
\begin{eqnarray}
\nonumber 
\dd s^2 &=&\beta^2 \dd t^2 -\gamma^{-2} z^{-4/(3+3w)}\dd r^2 \\
&&-r^{2(1+3w)/(3+3w)}t^{4/(3+3w)}\dd \Omega^2\; .
\label{eq:trans}
\end{eqnarray} 
This  is brought to a standard  form 
\begin{equation}
\dd s^2=\dd {\tilde t }^2-{\tilde t}^{4/(3+3w)}
\left(\dd {\tilde r}^2+{\tilde r}^2\dd \Omega^2\right)
\end{equation}
using the transformation
\begin{equation}
\label{eq:transrt}
\tilde t\equiv \beta t \quad {\rm and} \quad \tilde r = \beta^{-2/(3+3w)}r^{(1+3w)/(3+3w)}\; .
\end{equation}

\subsection{Equations in perturbed variable}

In order to describe deviations from the isotropic  FRW background,  we 
follow CY90 and introduce $A(z)$ and $B(z)$ according to\footnote{Note that 
the  $A$ is equal to $w$ times the function used by CY90.}
\begin{equation}
\label{eq:defab}
\xi\equiv z^{-2/(1+w)}e^{A} \; , \quad S\equiv z^{-2/(3+3w)}e^B \; . 
\end{equation}
From the definition of $\xi$ in (\ref{eq:def}), we get  $\mu t^2 =e^A$. Thus, to linear order $A$ can be identified as  the energy-density
constrast $\delta \mu/\mu$.  
The  equations (\ref{eq:phys1}) through (\ref{eq:phys4}), expressed   in terms of $A$ and $B$, 
are,
\begin{equation}
\ddot B=\frac{1}{3}\dot A+(1+w)\dot A \dot B -3 {\dot B}^2-
\left(\frac{1+3w}{1+w} \right)\dot B\; ,
\label{eq:com1}
\end{equation}
\begin{equation}
\dot B=\frac{1}{2}\dot A\left(1-\frac{w}{V^2}\right)+\frac{1}{3+3 w}
\left[e^{(w-1)A}-1\right]\; ,
\label{eq:com2}
\end{equation}

\begin{equation}
\label{eq:com3}
M=\left[\frac{1}{3}+(1+w)\dot B\right]e^{2B-(1+w)A}z^{2(1+3 w)/(3+3w)}\; ,
\end{equation}

\begin{equation}
\label{eq:com4}
M=\frac{1}{2}+\frac{3(1+w)^2}{4}e^{2B -2 w A} z^{2(1+3w)/(3+3 w)} \left[\dot B-\frac{2}{3(1+w)}\right]^2
\end{equation}
\begin{equation}
-\frac{9(1+w)^2}{2(1+3w)^2}\left[\dot B +\frac{1+3w}{3(1+w)}\right]^2e^{6B-2A}\; ,
\nonumber
\end{equation}
and
\begin{equation}
V=\left(\frac{1+3w}{\sqrt6}\right)z^{(1+3w)/(3+3w)}e^{(1-w)A-2B}\; .
\label{eq:com5}
\end{equation}

\subsection{Asymptotic solutions}
Linear perturbations describing small deviations from uniformity are
obtained for  $A\sim \delta \mu/\mu\ll 1$ and $\dot B \ll 1$.
The  linearized equations have two solutions. The first is  (CY90)
\begin{eqnarray}
A_1&=&-\frac{1+3w}{1+w} k z^{-2(1+3w)/(3+3w)}\; ,\\
B_1&=&B_\infty-k  z^{-2(1+3w)/(3+3w)} \; ,
\end{eqnarray}
where the constants $B_\infty$ and $k$ are related by 
\begin{equation}
k=\frac{3}{2}\frac{(1+w)(e^{-2B_\infty}-e^{4B_\infty})}{(1+3w)(5+3w)}\; .
\end{equation}
Expressed in terms of the FRW coordinates $\tilde r$ and $\tilde t$ given in (\ref{eq:transrt}), we see that $A_1\sim \delta \mu/\mu\propto {\tilde r}^{-2} {\tilde t}^{2(1+3w)/(3+3w)}$.  Therefore, the energy-density contrast, $\delta \mu/\mu$,  
is  ${\tilde r}^{-2}$, independent of $w$,  and has a temporal behavior 
identical to that  obtained for large scale perturbations in the standard linear perturbation theory 
in the comoving gauge (e.g. Peebles 1993). 
The second solution is 
\begin{equation}
\quad A_2(z)\propto z^{(1-w)/(1+w)} \quad {\rm and} \quad B_2(z)=\frac{A_2(z)}{6}\; .
\end{equation}
 According to (\ref{eq:transrt}), the limit  $\tilde r\rightarrow \infty$
 corresponds to $z\rightarrow \infty$ and $z\rightarrow 0$, respectively  for $w>-1/3$ and $-1<w<-1/3$. Therefore,
density perturbations approaching  zero as $\tilde r\rightarrow \infty$ are described by 
$A_1$ for  $w>-1/3$
and $A_2$ for $-1<w<-1/3$.  The mode $A_1$ grows with time, while $A_2$ 
decays.

\section{Shocks}
\label{sec:shock}

We will consider solutions involving shock discontinuities propagating in a medium of cold
pressure-less gas. We will assume that the heated matter behind the shock obeys the equation of state
$p=w\mu$. A  shock will be  specified by the ratio of the energy-densities on both of its sides.   The remaining hydrodynamical variables and the components of the  metric 
on both sides of the shock are then related by means of the jump conditions.  
These are usually derived by working first 
with Schwarzschild coordinates where the metric tensor is continuous and then 
transforming back to comoving coordinates (Bardeen 1965; Cahil \& Taub 1971).
 Quantities just ahead and behind the shock  are denoted by 
the $(+)$ and $(-)$ symbols, respectively. 
The general form of the jump conditions in comoving coordinates is (Cahil \&  Taub 1971),
\begin{eqnarray}
\label{eq:R} R_+&=&R_-\\
\label{eq:Vp} \Vp^2&=&\frac{(\wm\mum-\wp\mup)(\mum+\wp\mup)}{(\mum-\mup)(\mup+\wm\mum)}\\
\Vm^2&=&\frac{(\wm\mum-\wp\mup)(\mup+\wm\mum)}{(\mum-\mup)(\mum+\wp\mup)}\\
e^{2\num-2\nup}&=&\frac{(\mum+ \wp\mup)(\mup+\wp\mup)}{(\mup+\wm\mum)(\mum+\wm\mum)}\\
e^{2\lm-2\lp}&=&\frac{(\mup+ \wm\mum)(\mup+\wp\mup)}{(\mum+\wp\mup)(\mum+\wm\mum)}\\
\label{eq:Gamma}\Gamma_-&=&\frac{C+\Gamma_+}{1+C \Gamma_+}
\end{eqnarray}
where
\begin{equation}
\Gamma=e^{\lambda-\nu}\frac{\partial R/\partial t}{\partial R/\partial r} \quad {\rm and} \quad C=\frac{\Vp-\Vm}{1-\Vm\Vp}
\end{equation}
The condition (\ref{eq:R}) leads to the continuity of the 
self-similar variable $S$, i.e. $S_+=S_-$.
 Using the relations
\begin{eqnarray}
\nonumber R_t=rS_t&=&r\frac{\dd S}{\dd z}\frac{\dd z}{\dd t}=-z\dot S\; ,\\
\nonumber R_r=S+rS_r&=&S+r\frac{\dd S}{\dd z}\frac{\dd z}{\dd r}=S+\dot S\; .
\end{eqnarray}
we find that
\begin{equation}
\Gamma=\frac{-V(s) \dot S}{S+\dot S} \; .
\end{equation}
Therefore, the condition (\ref{eq:Gamma}) relates between $\dot S=z\dd S/\dd z$ on both sides of the shock.

\section{Numerical solutions} 
\label{sec:num}
We present here numerical solutions of the equations of motion for cases with and without shocks. 
The boundary conditions for all solutions obey 
the asymptotic form giving small density perturbations in the  limit of large FRW coordinate $\tilde r$, corresponding to $z\gg 1$ 
$w>-1/3 $ and to $z\ll 1$ for $w<-1/3$.

\subsection{Solutions without shocks}

\begin{figure}
\centering
\begin{sideways}
\mbox{\psfig{figure=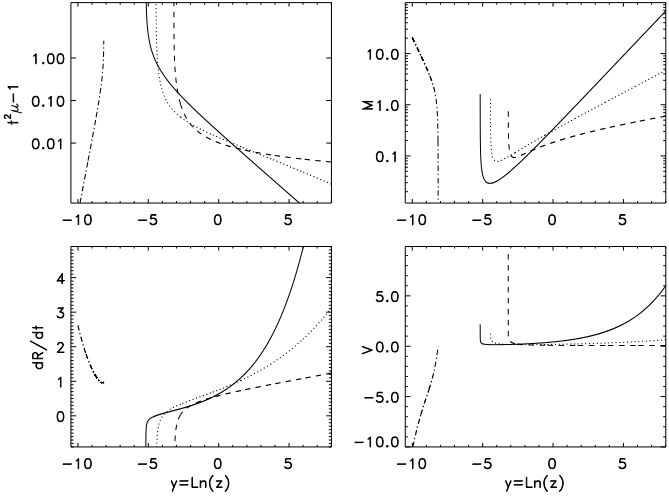,height=3.5in}}
\end{sideways}
\vspace{0.2cm}
\caption{Numerical solutions  of self-similar positive density perturbations in a FRW universe 
for  $w=0$ (solid lines), -0.2 (dotted), -0.3 (dashed) and  -0.6 (dash-dotted). 
 {\it Top-left:} the excess energy-density in units of the background value. 
 {Top-right:} the function $M=S/R$. 
 {\it Bottom-left:} the fluid ``velocity" $\dd R/\dd t=-z^2\dd S/\dd z$. 
 {\it Bottom-right:} the velocity $V$ of
 surfaces of constant $z$ relative to the fluid. 
 For $w=-0.6$, the  lines showing $M$, $\dd R/\dd t$ and $V$ are scaled down by  factors 
of $10^4$, 500 and 25, respectively.}
\label{fig:dwp}
\end{figure}

The case  $w>0$ has been considered extensively by CY90. Therefore, 
we only present  solutions for $-1<w<0$. An important difference between solutions with $w>0$ and $w<0$  is the following. 
For $w>0$, a sonic point at which $V=w^{-1/2} $ may exist. At this point the
term $(1-w/V^2)$ in equation (\ref{eq:com2}) vanishes, leaving $\dot A$ undetermined. 
Certain physical assumptions must then be made in order to complete the solution across the
sonic point (e.g. Ori \& Piran 1990;   CY90). For $w<0$, sonic points do not exist and, barring singularities,  
$\dot A$ is always determined by (\ref{eq:com2}).

Figure \ref{fig:dwp} shows results of numerical integration of a positive density perturbation for 
$w=0$ (solid lines), $w=-0.2$ (dotted),  $w=-0.3$ (dashed), and $w=-0.6$ (dash-dotted). 
The four panels show, from top-left in clockwise order, the contrast in the energy-density 
(i.e. $\mu/\bar \mu-1$ where $\bar \mu $ is the background energy density),  
the function $M=m/R$, the ``velocity" $\dd R /\dd t=-z^2\dd S/\dd z$, and the velocity $V$ 
of surfaces of constant $z$ relative to the fluid. 
Solutions with  $w>-1/3$ diverge at a finite value of $z$. This singular behavior of the solutions indicates the presence of black holes with 
apparent horizon at the  point with  $M=1/2$ and an event horizon at the point with $V=1$
(Carr \& Hawking 1974).
Comoving coordinates should allow a description of the 
physical variables  down to the
singularity at the center. For the solutions with $w>-1/3$, the finite value of $z$ at which the solutions diverge
represents  the shell having  a comoving coordinate $r$ and  arriving at
the center at  time $t=r/z$ (e.g. Landau \& Lifshitz 1975).
The  lines of $t^2 \mu-1 $ follow the asymptotic form  $z^{(1-w)/(1+w)}$
at $z\ll 1 $ for $w<-1/3$ and $z^{2(1+3w)/(3+3w)}$ at $z\gg 1$ for $w>-1/3$. 
For $w<-1/3$, the limit  $z\rightarrow 0 $ is obtained by 
 taking   the limit  $\tilde r\rightarrow \infty$ or/and $t\propto \rightarrow \infty$.
 Therefore, the self-similar solution for $w=-0.6$ describes a mode which decays with time and 
 approaches 
 zero at large FRW coordinate $\tilde r$, in contrast to the solutions shown for $w>-1/3$ which grow with time. 
In the Newtonian limit the quantity $\dd R /\dd t$ describes the velocity of a
shell relative to a central observer.  For $w>-1/3$, at sufficiently large $z$, 
$\dd R/\dd t$ is positive. As we decrease  $z$, $\dd R/\dd t$  changes signs, indicating the collapse of matter towards the center.

\subsection{Solutions with shocks}
\begin{figure}
\centering
\begin{sideways}
\mbox{\psfig{figure=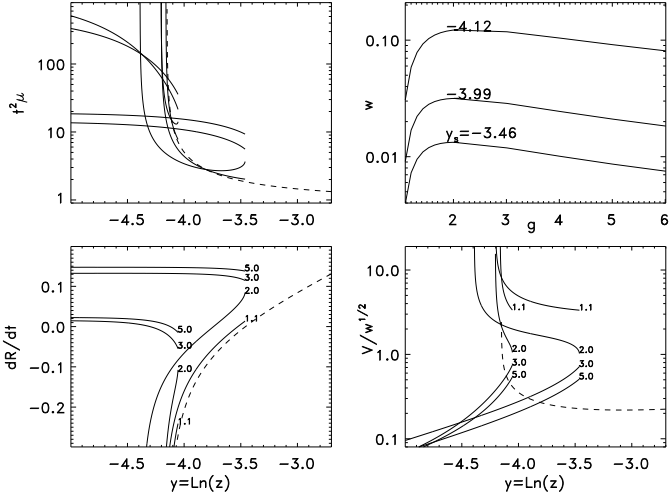,height=3.5in}}
\end{sideways}
\vspace{0.2cm}
\caption{Results of numerical integration of the equations in the presence of shocks
propagating from the center of a positive density excess. 
The dashed lines in the top-left, bottom-left and bottom-right panels 
correspond to the shock-less collapse of cold gas ($w=0$). The solid lines
in these panels 
represent  the physical variables in the shock-heated region.
Results are shown for shocks present  at $z_s=-3.46$ and $z_s=-4.05$; 
for $g=\mum/\mup=1.2$, 2, 3, and 5 at each $z_s$.
For these values of $z_s$ the shock-less solution has $\dd R/\dd t<0$.   
{\it Top-left:} the energy-density in units of the background density versus the variable $y=\ln z$. {\it Top-left:} the parameter $w$ in the shocked material as a function of the 
assumed $g$, for shocks at three values of $z_s$ as indicated in the figure. 
{\it Bottom-left:} the ``velocity" $\dd R /\dd t=-z^2\dd S/\dd z$.
{\it Bottom-right:}  the velocity $V$ of constant $z$ surfaces relative to 
the fluid, where the dashed line is the actual velocity of the collapsing cold gas ($w=0$), while 
the solid lines show $V/w^{1/2}$. }
\label{fig:dsh}
\end{figure}

We discuss now solutions describing self-similar perturbations 
in the presence  of shocks. The  cosmic background is assumed to be  made of cold pressure-less gas which is able to shock.
We focus first on shocks propagating from the center of 
a perturbation with positive energy-density excess in the cold gas.
  At the end of this section we present solutions for shocks propagating into the 
 uniform density background. 
The 
equations of motion (\ref{eq:com1}) and (\ref{eq:com2}) with $w=0$ are integrated inward from $z_1\gg 1$ 
with  initial conditions dictated by the linear solution. The integration gives tabulated values for the  variables $ A(z)$ and $B(z)$ 
from $z_1$ until  $z=z_s$ $(z_s<z_1)$, where a shock is assumed to exist.
Given the variables $A_+$ and $B_+$ just a head of the shock at $z_s$ 
we obtain the corresponding physical variables $\nu_+$, $\lambda_+$, $S_+$ and  $\xi_+$. 
Then, assuming  a value for  the ratio $g=\mum/\mup$ of the energy-densities behind and ahead of the shock, we 
use the jump conditions (\ref{eq:R}) through (\ref{eq:Gamma}) to obtain 
  $\nu_-$, $\lambda_-$, $S_-$ and  $\xi_-$ behind the shock. 
These  variables 
are then used as initial conditions 
in the numerical integration of the equations (\ref{eq:phys1}) and (\ref{eq:phys2}), from  $z_s$ down to smaller values.  This is done for several values $g$ and $z_s$. 
The condition (\ref{eq:Vp}) implies that shocks cannot exist at points 
with $V>1$. This is related to that fact that  $V=1$ marks  an event horizon 
inner to which matter must eventually hit the central singularity (Carr \& Hawking 1974).  
Results of the numerical integrations are shown in figure (\ref{fig:dsh}). 
The top-left panel plots the energy-density (in units of the baclground value),  $t^2\mu$. The dashed line represents the solution obtained for the shock-less collapse 
of cold gas ($w=0$).
The solid lines show $t^2\mu$ versus $z$  when shocks are assumed to exist at 
$\ln z_s=-3.46$ and $\ln z_s=-4.05$.  For these $z_s$, the shock-less solution 
has $\dd R/\dd t<0$ (the bottom panel to the left). For each value of $z_s$, lines representing solutions obtained for four values of $g=\mum/\mup$ are shown: $g=1.1$, 2, 3, and 5. 
At both $z_s$, the energy-density reaches a constant value as $z\rightarrow 0$
for  the two larger values of $g$ and diverges at a finite $z$ for the smaller values. 
Given $g$, the jump conditions determine  the pressure and energy-density
ahead of the shock and, therefore, the ratio $w=w_-=p_-/\mum$. 
The top-right panel plots lines of $w$ versus $g$ for three values of $y_s=\ln z_s$ as
indicates in the figure. 
The behavior of these lines can be easily understood by  a straightforward algebraic manipulation of  the jump condition (\ref{eq:Vp}) with $w_+=0$. 
For $g$ close to unity, $w$ increases linearly with $g-1$. In the limit $g\gg 1$
we get $w\approx \Vp^2/(1-\Vp^2)/g$ so that $p_-=w\mu_-$ which is independent of $g$.
Curves of $\dd R /\dd t$ are shown in the bottom-left panel. The numeric labels on each line indicate the corresponding value of
$g$. It is interesting that for  the regular solutions, $\dd R /\dd t$ converges to a positive constant  as $z\rightarrow 0$. This constant becomes smaller with 
decreasing $z_s$ and  we conjecture that it will approach zero as
$z_s$ gets very close to the event horizon marked by the point with $V=1$ in the dashed 
 curve (bottom panel to the left).  
The physical interpretation of this  is that outflows/winds from a central region are 
required for the formation of shocks. The intensity of these outflows determine
the location of the shock, $z_s$.  
 The bottom-right panel plots  the velocity $V$ versus $z$. The actual velocity is shown for the cold collapse (dashed) while the ratio $V/w^{1/2}$ is plotted for the other lines. For the regular  solutions, $V/w^{1/2}$ is less then unity throughout the shocked 
regions. The other solutions have $V/w^{1/2}>1$ over the range 
 from $z_s$ down to the finite $z$ marking the 
central singularity. Thus, in both cases no sonic points exist.   
Finally, the asymptotic behavior of the numerical  solutions  near the center is consistent with the
analysis of CC02.
For the  regular solutions, we have $V\rightarrow 0$ as $z\rightarrow 0$. The analysis of CC02 of this case gives  $S\propto z^{-1} $ and 
$\xi \propto z^{-2/(1+w)}$   which according to (\ref{eq:def})  implies $t^2\mu =const$ and $\dd R/\dd t =-z^2 \dd S/\dd z =const$.  Our solutions match this behavior extremely well.

\begin{figure}
\centering
\begin{sideways}
\mbox{\psfig{figure=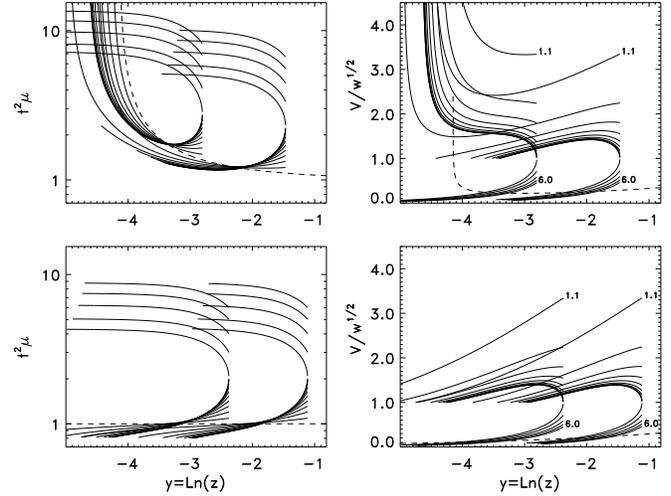,height=3.5in}}
\end{sideways}
\vspace{0.2cm}
\caption{ Comparison between solutions for shocks propagating in the uniform 
cosmic background (bottom panels) and in a positive density excess over the background (top). In all panels, dashed lines represent solutions without shocks, while the 
solid ones correspond  to shocks assumed to exist at two different positions.  The solid lines at each shock position correspond to several values of $g=\mum/\mup$ from $ g=1.1$, to 6 in equal steps
of 0.1.
Left panels show  lines of energy-density versus $y=\ln z$.   Right panels: 
show  $V/w^{1/2}$ for the solid lines
and the actual the velocity $V$ for the dashed line.   }
\label{fig:duni}
\end{figure}

It is interesting to compare between the solutions for shocks propagating in 
a positive density perturbation and in the uniform background. 
These solutions are represened, respectively, in the top and bottom rows of 
in figure (\ref{dif:duni}). The positive density 
solutions are taken from the output of the integration 
  used in figure (\ref{fig:dsh}) but 
for shocks assumed to exist 
at two points outside the region with $\dd R/\dd t<0$ 
(compare with figure (\ref{fig:dsh}).
This means that the shock exist at a
point where the cold fluid  is expanding. Physically this is possible if a central energy source is 
present.  
As is seen in the top panel to the right,   solid 
curves with intermediate $g$ hit  sonic points ($V=w^{1/2}$). At these points 
we stop the integration. Larger values of   $g$ yield  regular solutions without sonic points. 
For smaller $g$, the velocity $V$ changes direction and tends to infinity, indicating the
presence of a black hole. 
As $z_s$ is decreased, the solutions behave in a manner similar to what we have seen in 
figure (\ref{fig:dsh}). 
In the uniform background case, sonic points always exist for small $g$ and disappear for 
sufficiently large values.

\section{Discussion}           
\label{sec:disc}

In the standard cosmological model, the universe on large scales is nearly uniform so that
super-horizon perturbations  are well described by 
the general relativistic theory of linear perturbations 
  (e.g. Bardeen 1980).
  However, in view of resent observational data, it seems that GR might be  needed 
  for a full study of perturbations on small scales where strictly speaking the linearized
  equations are invalid. 
  These data suggest that the expansion of the universe is 
accelerating (e.g. Spergel et. al. 2003). It is thought that this acceleration is driven by 
a {\it dark-energy} component which, in its simplest incarnation, 
is described in terms of a fluid with pressure $p$ related to the energy density, $\mu$,  by 
$p=w \mu$ with $w\approx -1$ at the present cosmic epoch.   
The continuity equation, $\dot \mu=-(1+w) \mu$ implies that for $w=-1$ the density is 
 constant with time and, therefore, by Lorentz invariance, the 
fluid must be homogeneous. More generally, dark energy is modeled in terms of a scalar field 
which, under certain conditions,  could be  represented as a fluid with $w$ varying with time. 
Such a fluid is not ruled out by observations.  
If this is the case,  inhomogeneities in the dark energy component  could  develop. Since dark energy is inherently a relativistic entity, GR therefore should  be 
invoked for a thorough treatment of small scale cosmological perturbations. Seen in this perspective self-similar solutions might be relevant for the description of structure formation in the universe.  
However, the solutions are quite limited in their applicability in a realistic 
model for the growth of structure in the presence of dark energy. 
Putting the subdominant baryonic matter aside, the problem of structure formation 
requires tracing the evolution 
of the gravitationally coupled system of collisionless dark matter and
dark energy.   Our solutions do not take this coupling into account.  
It is unclear whether self-similar evolution of cosmological perturbations in a two fluid system is possible at all.
Further, we assume a strict equation of state of the form $p=w\mu$. This 
leads to an imaginary ``sound" speed, $(\delta p/\delta \mu)^{1/2}$ for $w<0$, while  a more realistic model for dark energy 
in terms of  scalar fields leads in general to a non-imaginary sound speed. 
Verertheless, because of the paucity of analytic solutions for time-dependent systems in GR,  a treatment of 
systems with restricted symmetry is a worthwhile effort. 
Analytic study of these systems could also  be  useful in testing and calibrating numerical 
techniques for solving the GR equations. 
\section{Acknowledgment} 
The author especially thanks  Amos Ori for many stimulating discussions
on general relativity and self-similarity.
He  also  thanks Ramy Brustein for useful discussions.

\protect



\end{document}